\documentclass[hyperref,letterpaper,floatfix,reprint,aps,pre,superscriptaddress,longbibliography]{revtex4-1}
\usepackage{graphicx}
\usepackage{amssymb,amsmath,amsfonts}
\usepackage[T1]{fontenc}
\usepackage{times}
\usepackage{hyperref}
\usepackage{array}
\usepackage{tabularx}
\usepackage{placeins}
\newcolumntype{M}{>{$\vcenter\bgroup\hbox\bgroup}c<{\egroup\egroup$}}

\usepackage{ifthen}
\usepackage{verbatim}

\usepackage{youngtab}
\usepackage{tikz}
\usetikzlibrary{arrows,shapes,positioning,calc,3d,shapes.geometric}
\usepackage{makecell}

\usepackage{color}

\definecolor{mbd}{HTML}{00274C}
\definecolor{mbl}{HTML}{587ABC}
\definecolor{mmaize}{HTML}{FFCB05}

\definecolor{gcb1}{cmyk}{0.35,0.07,0,0}
\definecolor{gcb2}{cmyk}{0.90,0.30,0,0}
\definecolor{gcb3}{cmyk}{0.30,0,0.45,0}
\definecolor{gcb4}{cmyk}{0.80,0,1.00,0}
\definecolor{gcb5}{cmyk}{0,0.40,0.25,0}
\definecolor{gcb6}{cmyk}{0.10,0.90,0.80,0}
\definecolor{gcb7}{cmyk}{0,0.25,0.50,0}
\definecolor{gcb8}{cmyk}{0,0.50,1.00,0}
\definecolor{gcb9}{cmyk}{0.20,0.25,0,0}
\definecolor{gcb10}{cmyk}{0.60,0.70,0,0}

\usepackage[normalem]{ulem}
\reversemarginpar

\begin{document}
\title{Topological Order in Densely Packed Anisotropic Colloids}
\author{William \surname{Zygmunt}}
\affiliation
{Department of Chemical Engineering, The University of Michigan,
Ann Arbor, MI 48109-2136, USA}
\author{Erin G.\ \surname{Teich}}
\affiliation
{Applied Physics, The University of Michigan,
Ann Arbor, MI 48109-1040, USA}
\author{Greg \surname{van Anders}}
\altaffiliation[Current Address: ]{Department of Physics, Engineering Physics, and Astronomy, Queen's University, 
Kingston Ontario, K7L 3N6, Canada, gva@queensu.ca}
\affiliation
{Department of Chemical Engineering, The University of Michigan,
Ann Arbor, MI 48109-2136, USA}
\affiliation
{Applied Physics, The University of Michigan,
Ann Arbor, MI 48109-1040, USA}
\affiliation
{Department of Physics, The University of Michigan,
Ann Arbor, MI 48109-1040, USA}
\author{Sharon C.\ \surname{Glotzer}}
\affiliation
{Department of Chemical Engineering, The University of Michigan,
Ann Arbor, MI 48109-2136, USA}
\affiliation
{Applied Physics, The University of Michigan,
Ann Arbor, MI 48109-1040, USA}
\affiliation
{Department of Physics, The University of Michigan,
Ann Arbor, MI 48109-1040, USA}
\affiliation
{Department of Materials Science and Engineering, University of Michigan, Ann
Arbor, MI 48109-2136, USA}
\begin{abstract}
The existence of topological order is frequently associated with strongly
coupled quantum matter. Here, we demonstrate the existence of topological phases
in classical systems of densely packed, hard, anisotropic polyhedrally shaped
colloidal particles. We show that previously reported transitions in dense
packings lead to the existence of topologically ordered thermodynamic phases,
which we show are stable away from the dense packing limit.  Our work expands
the library of known topological phases, whose experimental realization could
provide new means for constructing plasmonic materials that are robust in
the presence of fluctuations.
\end{abstract}
\maketitle
\parskip 0pt

Topological phases are exotic states of matter that are typically associated with strongly interacting quantum systems, in which topological protection stabilizes certain physical behaviors against
environmental perturbations \cite{topologicalmemory}. In quantum systems, protection of this type can be invaluable in applications for which coherence is crucial.
In a similar spirit, many applications for classical soft matter systems of colloidal nanoparticles would benefit from topological order in the presence of environmental perturbations. In colloidal systems, entropic effects are important \cite{entint,ordviaent, dfamilya} and typical interaction strengths are on the order of the thermal scale. Indeed, recent work has shown that thermal fluctuations in soft systems can, in a variety of contexts, drive structural reconfiguration \cite{gang11,shpphphshp,quasi2d2stepssxfm,shapesolidsolid}, an important feature of functional nanomaterials.  However, for other applications, the preservation of structural order against thermal fluctuations is vital.  If soft matter was topologically ordered, it could provide for building robust soft matter structures.

Recent work \cite{kanelubensky} has shown that topological states can exist in specialized classical mechanical systems. The topological states that occur in those systems are expressed in terms of 
Witten indices, whose existence relies on the spectrum of excitations in nearly-isostatic lattices. To date, no topologically protected phases have been reported in other classical systems.

Here, we show that the point-set topology of contacts that distinguishes
structures of hard colloids at infinite pressure (aka ``putative densest
packings'') \cite{dfamilyp} leads to the existence of topologically distinct phases. We prove analytically that, in general, topologically distinct putative densest packings lead to the existence of associated thermodynamic phases away from infinite pressure. We demonstrate numerically that topological order persists at finite pressure. Surprisingly, we find that thermodynamic phases that are topologically protected at the highest possible packing densities preserve near-perfect topological order at packing densities sufficiently low that topological protection need not persist. 

Our approach provides a general framework for investigating and classifying the structure of thermodynamic systems of hard colloids near the dense packing limit. 
The topological order we observe is of a strikingly different origin -- and
consequently, has different properties -- than topological states in quantum matter. Ref.\ \cite{kanelubensky}  showed that the existence of topological phases is not uniquely the preserve of strongly interacting quantum matter, and our results raise the possibility that topological order is a widespread phenomenon in classical systems.

\begin{figure}
  \includegraphics[width=8.6cm]{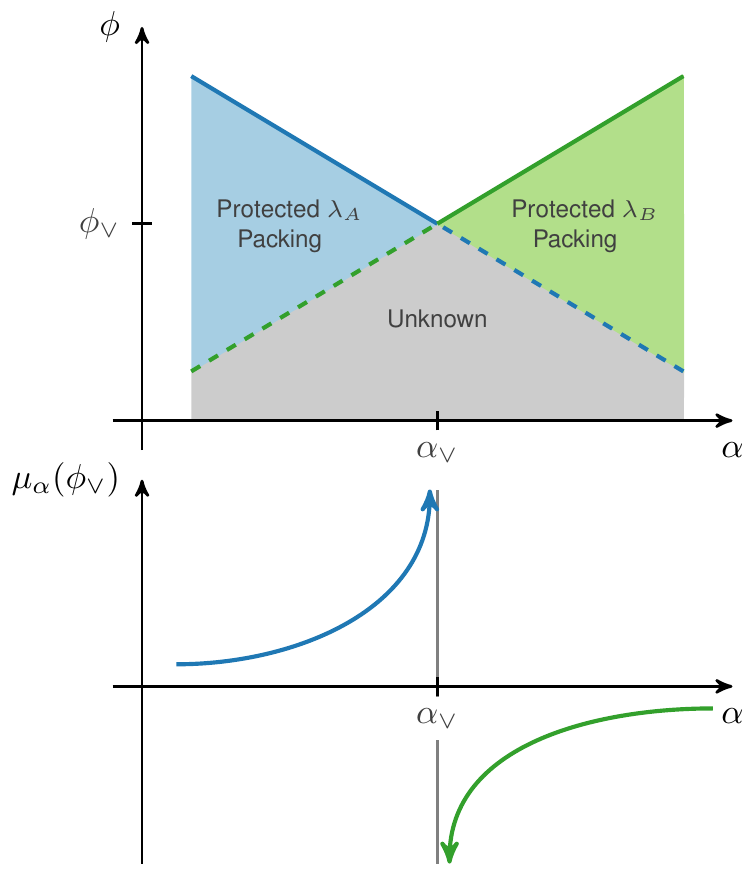}
  \caption{Surfaces (solid blue and green lines) of maximal packing density ($\phi_d$) as a function of particle shape ($\alpha$) have been shown \cite{dfamilyp}, in general, 
  to exhibit non-analytic behavior at the point $(\alpha_\vee,\phi_\vee )$ that is associated with a change in the topology of contacts between adjacent particles. 
  Within the blue and green triangular regions, dense packings exhibit topological distinction by particle contacts. However, within the gray 
  region bounded above by the dashed blue and green lines, it is unknown, in general, whether topological contact types persist (top panel). We 
  show (see text) that there is a first-order transition, indicated by the divergence of the so-called alchemical potential \cite{digitalalchemy} near the valley packing discontinuity (bottom panel).}
  \label{protected_phase}
\end{figure}


Ref.\ \cite{dfamilyp} showed that, for families of hard anisotropic shapes at infinite pressure, continuous deformations of particle shape result in continuous changes in putative densest packings. It was also shown that non-analytic behavior in curves, surfaces, or hypersurfaces $\phi_d(\alpha_i)$ of maximal packing density $\phi_d$ as a function of particle shape $\alpha_i$ occur if and only if there is a change in the point-set topology of contacts between particles in the dense packing structure; Fig.\ \ref{protected_phase} illustrates this point. We argue that, generically, this leads to distinct thermodynamic phases away from infinite pressure. Since pressure is defined in units of $k_{B}T$, we can think of the infinite pressure limit as a sort of zero temperature limit of the system, which may be a more useful way to think about these hard particle systems in reference to other works involving topological order. 

We make our argument using the framework of digital alchemy
\cite{digitalalchemy}. Digital alchemy extends the traditional thermodynamic
ensemble for particle self-assembly through consideration of thermodynamically
conjugate variables (termed ``alchemical potentials'' $\mu$) coupled to changes in particle attributes. Here, we consider changes in particle shape as the alchemical variable for the packings in \cite{dfamilyp} and the relevant alchemical potential is given by
\begin{equation} \label{alch_pot}
  \mu_\alpha = \frac{1}{N} \frac{\partial F}{\partial \alpha}\; ,
\end{equation}
where $N$ is the number of particles in the system, $F$ is the free energy and $\alpha$ is a shape parameter. To understand the phase behavior of dense suspensions of anisotropic colloids away from infinite pressure, it is convenient to study the alchemical potential in the vicinity of the intersection of two packing curves, where there is a change in the topology of particle contacts ($\alpha_\vee,\phi_\vee$ in Fig.\ \ref{protected_phase}). There, we can Taylor expand Eq.\ \eqref{alch_pot} for some $\phi < \phi_\vee$, as detailed in the Supplementary Material, to give $\mu_\alpha$ to leading order in the dimensionless pressure $P^*$
\begin{equation} \label{muapprox}
  \mu_\alpha \approx \pm 
  \frac{1}{2\beta}
  \left|\frac{\partial\phi_d}{\partial\alpha}\right|
  \left(\frac{P^*(\phi_d)}{\phi_d^2}+\frac{P^*(\phi_\vee)}{\phi_\vee^2}
  \right) \; ,
\end{equation}
where $\beta$ is inverse temperature, $\phi_d$ is maximal packing density, and the overall sign is determined by whether we carry out the expansion for $\alpha$ greater or less than $\alpha_\vee$. $P^*(\phi_\vee)$ diverges by definition as it represents the infinite pressure surface of maximal packing density; this means that $\mu_\alpha$ diverges near $\phi_\vee$, with a sign that changes across $\alpha_\vee$. Since $\mu_\alpha$ is a first derivative of the free energy $F$, this divergence in $\mu_\alpha$ indicates a discontinuity in $F$, which in turn indicates a thermodynamic phase transition. This transition exists solely because of the non-analytic behavior of the dense packing surface, which reflects the topology of contacts among densely packed particles.

Next we consider what happens below maximum packing density. We consider packings $\lambda_i$ where $i  \in \{A,B\}$; $\lambda_A$ and $\lambda_B$ are on either side
of the phase transition shown schematically in Fig.\ \ref{protected_phase}. To distinguish between these packings we construct an order parameter that takes advantage of the 
way $\lambda_A$ and $\lambda_B$  are defined topologically.  Ref. \cite{dfamilyp} defines each packing according to the types of contacts (face-face, face-vertex, face-edge,
vertex-vertex, vertex-edge and edge-edge) shared between adjacent particles. These contacts map to a set of intersection equations that mathematically describe each contact by 
relating particle shape parameters to the geometry of the unit cell of the packing described by vectors for the lattice and particle(s) within the unit cell. Each packing $\lambda_i$ 
has $K_i$ unique (meaning unshared with the other packing) intersection inequalities $\bigl|C_{i,k}\bigr| \geq 0$ (where $k = 1, 2, \ldots, K_i$) that define the packing. When the shape parameters and unit cell geometry correspond to densest packing, all the $C_{i,k}=0$. If the geometry of the unit cell does not correspond to the densest packing, then some $\bigl|C_{i,k}\bigr| \geq 0$. 
Changes in unit cell geometry (while particle shape is fixed) effectively 
 provide a means of measuring changes to particle contact; 
 at lower packing densities, then, the saturation or near-saturation of the intersection inequalities (i.e. all $C_{i,k} \approx 0$) would imply that particle contacts have (through thermal fluctuation) remained approximately equal to the particle contacts at infinite pressure, preserving unit cell geometry and topological order.

We now define an order parameter of the form $\theta_{ij}$ where $i$ represents a stable or metastable thermodynamic phase that is putatively isostructural with $\lambda_i$ \footnote{Note that if subsequent evaluation of the order parameter indicates the state is not isostructural with $\lambda_i$ it implies those structures are thermodynamically unstable} and $j$ is the packing type of the set of intersection equations against which the state will be evaluated. For example, $\theta_{AA}$ is defined as the evaluation of a packing type $\lambda_A$ in its own intersection equations (those of $\lambda_A$), and it evaluates to unity. Conversely, $\theta_{AB}$ (pertaining to the same packing type $\lambda_A$ evaluated in the intersection equations of $\lambda_{B}$) evaluates to zero.

First, we describe the variables that we will use to construct $\theta_{ij}$. The packing ${\lambda_i}$ is a function of particle shape $\alpha$ and packing density $\phi \leq \phi_d$ by definition. 
We define $\xi_{ij}$ (a function of particle shape $\alpha$ and packing density $\phi$) to reflect the evaluation of packing $\lambda_i$ in the intersection equations of packing type $\lambda_j$ as
\begin{equation} \label{OP_thermal}
	\xi_{ij}(\phi, \alpha) = e^{-\frac{1}{K_j}{\sum\limits_{k=1}^{K_j}\bigl|{C_{j,k}(\lambda_{i}(\phi,\alpha))}\bigr|}}.
 \end{equation}
When $\phi = \phi_d$, $\xi_{ij}$ describes the satisfaction of $\lambda_j$'s intersection equations by $\lambda_i$ at its maximum packing density, lying on its putative densest packing surface. We thus denote this special case by
 \begin{equation}
	\xi^\text{ideal}_{ij} = \xi_{ij}(\phi_{d}, \alpha).
 \end{equation}
 
\begin{figure*}
  \includegraphics[width=18cm]{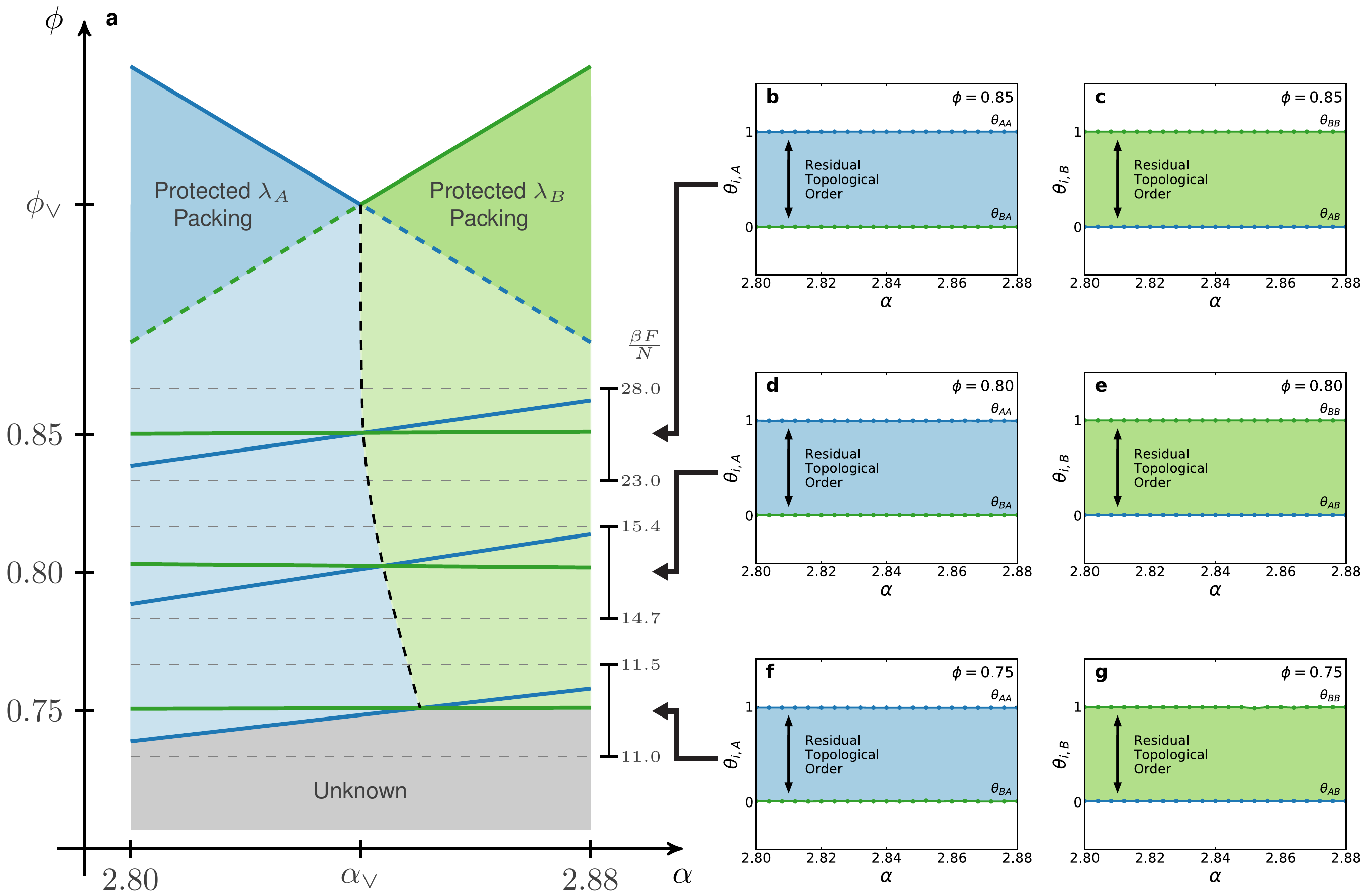}
  \caption{Panel (a) shows the curves of maximal packing density at $\phi_\vee$, outlining the two protected packing regions, where packing $\lambda_A$ is blue and packing $\lambda_B$ is green.      Lower curves indicate computed free energies at three packing densities (0.85, 0.80, 0.75). Darker shaded colors indicate protected regions, while lighter shaded colors indicate free energy preferred regions of the ($\alpha$, $\phi$) phase diagram. The gray region is a region where the preferred phase is unknown. Panels (b, d, f) indicate topological order evaluated using the intersection equations for $\lambda_A$ and panels (c, e, g) indicate the same using the intersection equations for $\lambda_B$. The dotted black line roughly demarcates boundaries between thermodynamically preferred packings as a function of packing density, and is meant to guide the eye.
  \label{52_58_all_parts_new_version}
  }
\end{figure*}

To construct a generalized order parameter for a set of two adjacent packings $\lambda_A$ and $\lambda_B$, we may then compute four quantities ($\xi_{AA},\xi_{AB},\xi_{BA},\xi_{BB}$, which consider all four evaluation types in $i  \in \{A,B\}$ and $j  \in \{A,B\}$), which we use to build vectors that represent coordinates in the $[\xi_{iA},\xi_{iB}]$ plane
 \begin{equation} \label{ivecs}
 	\bold{D_{A}} = \bigg [\xi_{AA}, \xi_{AB} \bigg ]   \;\:,\:\;\bold{D_{B}} = \bigg  [\xi_{BA}, \xi_{BB} \bigg ].
 \end{equation} 
 Similarly, we construct vectors to represent the maximum density packings
   \begin{equation} \label{ideal_vecs}
 	\bold{D^\textbf{ideal}_\textbf{A}} = \bigg [\xi^\text{ideal}_{AA}, \xi^\text{ideal}_{AB} \bigg ]  \;\:,\:\; \bold{D^\textbf{ideal}_\textbf{B}} = \bigg  [\xi^\text{ideal}_{BA}, \xi^\text{ideal}_{BB} \bigg ].
 \end{equation} 
 The distance between the ideal structures in the $[\xi_{iA},\xi_{iB}]$ plane is
 \begin{equation} \label{ideal_dist}
 	\bold{D^\textbf{ideal}_\textbf{AB}} = \sqrt{(\bold{D^\textbf{ideal}_{A}}-\bold{D^\textbf{ideal}_{B}}) \cdot (\bold{D^\textbf{ideal}_{A}}-\bold{D^\textbf{ideal}_{B}})}.
 \end{equation} 
Finally, we define $\theta_{ij}$ to distinguish the topology of the two packings, making a generalized expression for any packing type $\lambda_i$ evaluated in the intersection equations of a packing type $\lambda_j$
 \begin{equation} \label{OP_ij}
 	\theta_{ij} = 1 - \frac{\sqrt{(\bold{D_{i}}-\bold{D^\textbf{ideal}_{j}}) \cdot (\bold{D_{i}}-\bold{D^\textbf{ideal}_{j}})}}{\bold{D^\textbf{ideal}_\textbf{AB}}}.
 \end{equation}
The evaluation of Eq.\ \eqref{OP_ij} on an example system (described below) is shown in Fig.\ \ref{52_58_all_parts_new_version}(b), where both $\theta_{AA}$ and $\theta_{BA}$ are plotted, showing that when $\lambda_A$ is evaluated in the intersection equations of $\lambda_A$, the order parameter is unity. When $\lambda_B$ is evaluated in the intersection equations of $\lambda_A$, the order parameter is zero.

\begin{figure}
  \includegraphics[width=8.5cm]{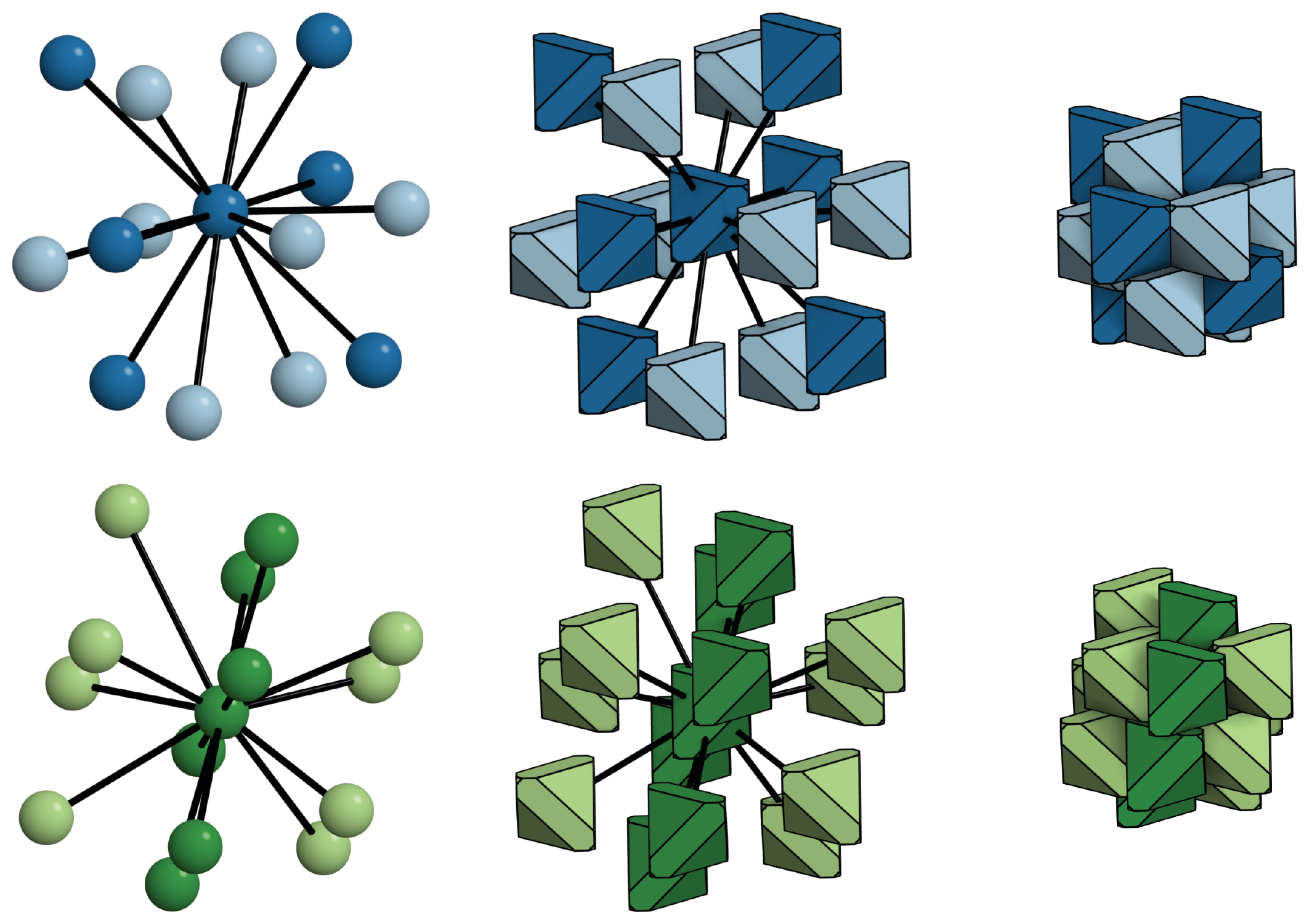}
  \caption{Example dense packing structures (where packing $\lambda_A$ is blue and packing $\lambda_B$ is green) of anisotropic
    shapes, including ``exploded'' views that show the location and orientation
    of neighboring particles, and densely packed units.
  \label{shapes5258}
  }
\end{figure}

For concreteness, we considered packings in the two-parameter family of triangle invariant polyhedral shapes, $\Delta_{323}$, reported in Ref.\ \cite{dfamilyp}. This family of shapes includes three Platonic solids (tetrahedron, octahedron, cube) and truncations thereof. Those authors showed the existence of 75 topologically distinct two-particle dense packings of polyhedra in this family.  As one example, we studied the boundary between phases labelled `52' and `58' in Ref.\ \cite{dfamilyp}, which we refer to hereafter as $\lambda_{A}$ and $\lambda_{B}$, respectively (the difference in topology of these packings is shown in Fig.\ \ref{shapes5258}). We studied a range of constituent particles with shape variables $\alpha_a=(2.80,2.88)$ and $\alpha_c=1.52$.  We set one $\alpha$ to be constant and moved along the axis of the other $\alpha$.

We initialized systems of 1024 identical particles with shape variables $( \alpha_a, \alpha_c )$ (denoted by $(u,v)$ in the notation of Ref. [10]) in both $\lambda_{A}$ and $\lambda_{B}$ at various densities. Particle positions and orientations are well defined for initialization in Ref. [10]. We sampled systems in the isochoric ensemble using the hard particle Monte Carlo (HPMC) \cite{Anderson2015a} extension of the simulation toolkit HOOMD-blue \cite{hoomdblue, Glaser2015}. Although the volume remained fixed, box shear and aspect ratio moves were allowed, and move sizes were tuned such that acceptance ratios were approximately 0.3. We calculated pressure during these simulations via the scaled distribution function \cite{hexatic}, whose measurement is implemented in HPMC \cite{hpmcplug}. Ensemble averages were taken over five replicates and five snapshots per replicate simulation, where each simulation snapshot is separated by $1.0 \times 10^{6}$ MC timesteps, well beyond the calculated autocorrelation time of the system pressure. For each data point, we constructed a system in the ideal putative densest packing structure and then expanded this structure to the target packing density.

Free energies were computed via the Frenkel-Ladd \cite{frenkelladd,amirtetphdiag} method. The Einstein crystals for these simulations were the same packings described above, with an expansion performed down to the desired packing density achieved at the beginning of the simulation. An external force field $\Lambda$ tethered particles to their crystal sites with a spring constant of $k = \exp(25)$ which has units of $k_{B}T$. We fixed length units by taking particles to have unit volume. Every $1.4 \times 10^5$ timesteps, $k$ was decreased until it was eventually 0; each time $k$ was changed, move sizes were tuned,  $1.0 \times 10^5$ timesteps were run for equilibration and the lattice energy was calculated in the remaining $4.0 \times 10^4$ timesteps.

Fig.\ \ref{52_58_all_parts_new_version}(a) shows the curves of maximal packing density $\phi_d$ for each packing type, indicating the protected regions in darker shading under them. Below that are curves of free energy as a function of both packing and shape at various packing densities $\phi$ well below the maximum packing density $\phi_d$. The plots of free energy show that even at densities well below this value two phases persist up to some crossing. The location of this crossing at densities below $\phi_\vee$ need not be at $\alpha_\vee$, and we find that it does deviate from $\alpha_\vee$ at lower packing density. The free energy preferred regions are colored in lighter shades of the protected regions and to verify that the existence of the two phases at finite pressure arises from the topology of particle packing, we compute the relevant order parameters from Eq.\ \eqref{OP_ij} for each structure. This calculation is performed by extracting the unit cells of the thermalized packings of $\lambda_A$ and $\lambda_B$ at a packing density $\phi_d$.
The unit cell extraction technique is outlined in the Supplemental Material. 

We found that at densities well below $\phi_\vee$, phases identified by the free energy calculation correspond to phases that differ in the topology of particle contacts measured through the order parameters $\theta_{ij}$. In Fig.\ \ref{52_58_all_parts_new_version}(b,d,f) we evaluated the order parameter $\theta_{iA}$ on structures $\lambda_{i}$ where $i  \in \{A,B\}$
and found that over a range of packing densities, $\theta_{iA}$ evaluates to near unity on $\lambda_{A}$ and vanishes on $\lambda_{B}$. Conversely, in Fig.\ \ref{52_58_all_parts_new_version}(c,e,g) we evaluated the order parameter $\theta_{iB}$ on $\lambda_{A}$ and $\lambda_{B}$ and found that over a range of packing densities, $\theta_{iB}$ evaluates to near unity on $\lambda_{B}$ and vanishes on $\lambda_{A}$. These results indicate that the phases can be identified by the topology of the related putative densest packings, and possess residual topological order, or order that matches the order of a topological state at a packing density where topological protection has not been proven to exist. The residual topological order we observe in Fig.\
\ref{52_58_all_parts_new_version} suggests that crystal structures present in densely packed colloidal suspensions maintain a topologically consistent set of contacts between particles at densities where other competing contact topologies could exist, but are unlikely to do so due to the existence of a more thermodynamically favorable topological state.
A second system (also in the densest packing landscape) was picked arbitrarily for evaluation in order to reinforce the results shown here; the similar results for that system can be found in the Supplementary Material.

The topologically distinct phases of dense suspensions of anisotropic colloids that we find here are dissimilar to topological phases in quantum matter in almost all respects, except in their stability against perturbations. For instance, whereas the topological entropy of ground-state degeneracy that arises from entanglement is important in quantum systems \cite{topologicalentropy}, in our systems, instead, shape entropy \cite{entint} quantifies ground state degeneracy. Moreover, whereas the geometric topology that underlies topological order in quantum systems allows a considerable mathematical apparatus to be brought to bear in understanding those states, the point-set topology that underlies the classical, topological order we identify here is more limited. Nevertheless, despite the rudimentary form of the topological order reported here, colloidal systems remain robust against perturbation, since they persist even at lower packing densities where topological protection is no longer required. This robust persistence would be a key desirable feature for applications in regimes away from the infinite pressure limit. Moreover, because the form of topological order is more rudimentary, previous work \cite{dfamilyp} demonstrating that topological features (such as particle contact types between faces, edges and vertices) generically distinguish phases of densely packed colloids suggests that this form of topological order is widespread in colloidal systems \cite{dijkstranonconvex,escobedo,dijkstrasuperballs,trunctet,dijkstratcube}.

To leverage this topological order in experiment we note that though our order parameters are based on contact types that nominally arise at infinite pressure, we showed that topological order persists at finite pressure. This finding is potentially useful in constructing plasmonic materials that have robust response in the presence of thermal fluctuations, changes in particle shape \cite{epp} or the behavior of stabilizing ligands \cite{Waltmann2017,Waltmann2018}. It is known that the plasmonic response of systems of anisotropic nanoparticles depends strongly on the type of contacts between nanoparticles \cite{taoetal}. We find that the topology of contacts between anisotropic nanoparticles is stable over a broad range of packing densities. When situated in the context of the zoo of distinct sets of contact types that has been shown to exist \cite{dfamilyp} in families of anisotropic nanoparticles and the variety of synthesis techniques that can readily produce such particles in the laboratory \cite{skrabalak,geissleryang,huang2012facedependent,liao2013polyhedralgold}, our work points to potential avenues for creating nanomaterials with a diversity of robust forms of plasmonic response.

\section{Acknowledgments}

Research for this publication was conducted with Government support under contract FA9550-11-C-0028 and awarded by the Department of Defense, Air Force Office of Scientific Research, National Defense Science and Engineering Graduate (NDSEG) Fellowship, 32 CFR 168a, awarded to W.E.Z. E.G.T acknowledges support from the National Science Foundation Graduate Research Fellowship Grant DGE 1256260 and a Blue Waters Graduate Fellowship. This research is part of the Blue Waters sustained petascale computing project, which is supported by the National Science Foundation (awards OCI-0725070 and ACI-1238993) and the state of Illinois. Blue Waters is a joint effort of the University of Illinois at Urbana-Champaign and its National Center for Supercomputing Applications. This work was partially supported by a Simons Investigator award from the Simons Foundation to Sharon Glotzer. Computational resources and services supported by Advanced Research Computing at the University of Michigan, Ann Arbor. We thank D. Klotsa and E. R. Chen for access to data pertinent to this study and helpful discussion of that data. We also  thank J. Dshemuchadse and B. VanSaders for helpful discussions and software contributions.

\bibliography{gmaster}{}
\bibliographystyle{apsrev4-1}

\end{document}